\documentclass[AMA,STIX1COL, doublespace]{WileyNJD-v2}
\articletype{Article Type}%

\received{26 April 2016}
\revised{6 June 2016}
\accepted{6 June 2016}

\DeclareMathOperator*{\argmin}{arg\,min}

\begin{document}

\title{Is Group Testing Ready for Prime-Time in Disease Identification?}

\author[1]{Gregory Haber}
\author[2]{Yaakov Malinovsky}
\author[1]{Paul S. Albert*}

\address[1]{\orgdiv{{Biostatistics Branch, Division of Cancer Epidemiology and Genetics,},
    \orgname{National Cancer Institute}, \orgaddress{Rockville, MD 20852}}}

\address[2]{\orgdiv{{Department of Mathematics and Statistics},
    \orgname{University of Maryland, Baltimore County}, \orgaddress{Baltimore, MD 21250}}}

\corres{*Biostatistics Branch, Division of Cancer Epidemiology and Genetics, National Cancer Institute, Rockville, MD 20852}

\abstract[Summary]{ Large scale disease screening is a complicated process in which high costs must
be balanced against pressing public health needs. When the goal is screening for infectious disease,
one approach is group testing in which samples are initially tested in pools
and individual samples are retested only if the initial pooled test was positive.  Intuitively, if
the prevalence of infection is small, this could result in a large reduction of the
total number of tests required. Despite this, the use of group testing in medical studies
has been limited, largely due to skepticism about the impact of pooling on the accuracy of a given
assay. While there is a large body of research addressing the issue of testing errors in group
testing studies, it is customary to assume that the misclassification parameters are known from an
external population and/or that the values do not change with the group size. Both of these
assumptions are highly questionable for many medical practitioners considering group testing in their study
design. In this article, we explore how the failure of these assumptions might impact the efficacy of a
group testing design and, consequently, whether group testing is currently feasible for medical
screening. Specifically, we look at how incorrect assumptions about the sensitivity
function at the design stage can lead to poor estimation of a procedure’s overall sensitivity and
expected number of tests. Furthermore, if a validation study is used to estimate the pooled
misclassification parameters of a given assay, we show that the sample sizes required are so large
as to be prohibitive in all but the largest screening programs.  }


\maketitle

\section{Introduction} \label{Intro} Developing design strategies to reduce
study expense is an important job for the practicing biostatistician. In many
settings, measuring biomarkers can be expensive, and design strategies for
reducing these costs are needed. In 1943, Dorfman\cite{dorfman1943} proposed
a simple method to make the testing of syphilis feasible in recruits for the
U.S. Army. This simple design suggested testing a grouped collection of k
samples, and only testing individual samples if the combined sample is
positive. Intuitively, this design could provide a tremendous cost reduction
in terms of the required number of tests if the disease prevalence is small.

There has been a vast amount of research in group testing since the original
Dorfman paper. A majority of this work can be divided into either using group
testing for disease screening or for prevalence estimation. Particularly for
the application of group testing to screening, there has been lots of work
not only in the statistics literature, but also in computer science and
applied mathematics (\citep[see,][among others]{aldridge2019, bar-lev2010,
zhigljavsky2003a, macula1999, mcmahan2012}). This is because optimality of
group testing algorithms often involves choosing a particular set of group
sizes that minimizes the expected number of tests required to identify all
cases in a population of individuals. Deriving optimal designs often involves
the development of complex algorithms that rely on dynamic programming,
mathematical techniques that are usually applied in areas of applied
mathematics and computer science, and less often in statistics. There has
also been extensive work in prevalence estimation, where participants are
tested in groups with no re-testing at the individual level. This article
will focus on the use of group testing in disease screening.

Although methodological research in this area has expanded, we believe that
there has been limited use of these designs as they were originally
formulated in the biomedical sciences for disease screening. Some notable
exceptions do exist, such as the screening of donated blood for HIV and
hepatitis.\cite{stramer2009} More recently, group testing has received
increasing attention for SARS-CoV-2 screening. However, the successful
implementation of group testing in this area remains unclear at this
point.\cite{yelin2020, abdalhamid2020, litvak2020} We have seen a reluctance
by our epidemiological collaborators to use these designs in large scale
studies. In part this is due to our laboratory and epidemiology colleagues
not being aware of the advantages of the group testing methodology. However,
most often, scientists are afraid that combining different participants in a
single sample will decrease the sensitivity of an assay, thereby increasing
the likelihood of a false negative on a grouped test. Furthermore, it is
perceived that in much of the group testing literature unreasonable
assumptions have been made that, in many cases, favor the use of group
testing procedures over single testing. The goal of this current article is
to provide a balanced view of the research in this area and to provide
suggestions for evaluating its feasibility in practical settings.

There are a number of issues that have caused confusion and have made comparisons with individual
testing difficult. These include:

\begin{itemize}
    \item[i.] Questions over how to choose a design that appropriately accounts for misclassification.
    \item[ii.] Assumptions of non-differential misclassification, that is, that the testing errors do not
        change with the group size.
    \item[iii.] Assumptions that sensitivity and specificity values are known a priori from external
        sources and can be readily applied to the question at hand.
\end{itemize}

A careful comparison of group testing with individual testing that takes into
account these issues is important in deciding the situations where group
testing should be used for disease identification in the biosciences.
Although aspects of these issues have been addressed in the statistics literature, there has not been a careful
statistical examination of these issues in totality. In
what follows, we introduce several case studies that are representative of
the types of problems in which group testing is appealing to researchers, but
current limitations raise questions about or prevent its use. This is
followed by a full discussion of each of the above issues. We then present
numerical comparisons to examine the impact of incorrect assumptions
regarding the misclassification parameters on a screening procedure and the
feasibility of using a validation study to estimate these values.

\section{Case studies}

\subsection{Population based screening for SARS-CoV-2 infection}
Coronavirus disease 2019 (COVID-19) caused by the severe acute respiratory syndrome coronavirus 2
(SARS-CoV-2) was first identified in Wuhan, China in late 2019 and is rapidly
spreading worldwide with dramatic impacts on the healthcare and economic landscapes. In the United
States, shortages of testing reagents hinder the ability to carry out sufficient screening for
infection with SARS-CoV-2 which may ultimately threaten the ability of public health officials to
adequately control the spread of the virus. The need for large scale screening, coupled with
scarce testing resources, make this an ideal scenario for implementing group testing to reduce the
number of tests required to carry out a screening program.

\subsection{Large scale screening for HIV viral load}
Monitoring of viral load in individuals diagnosed with HIV is important for determining treatment
failure and making informed treatment decisions. Current World Health Organization guidelines
recommend viral load testing at 6 and 12 months following initiation of antiretroviral therapy (ART), and
annually thereafter.\cite{worldhealthorganization2016}  These tests are meant to determine if
an individuals viral-load has surpassed a given threshold, potentially indicating treatment failure. Current recommendations
define treatment failure as a test measuring a viral load higher than 1000 copies/ml.
These recommendations have proven to be cost prohibitive, and continue to be
impractical in many low resource settings. For example, in Malawi, a country with over 1 million
people living with HIV and over 600,000 taking ART,\cite{kanyerere2016} the annual burden of sufficient viral load
monitoring is enormous. In this context, group testing has regularly been considered as a cost
saving measure, however concerns over false negative rates are
common.\cite{rowley2014,elbouzidi2014, tilghman2011}

\subsection{Stratified cancer screening} Cervical cancer is currently the 4th
most common cancer among women,\cite{hernandez-lopez2019} highlighting the
need for cheap and effective clinical screening. The most effective indicator
of cervical cancer risk is HPV infection, however only a small number of HPV
positive women will go on to develop cervical cancer. To minimize unnecessary
and invasive follow-up procedures such as colposcopy, it is necessary to
develop better tests for triaging HPV infections in order to identify those
at greater risk of progressing to cancer. One promising method is methylation
testing which captures the methylation of HPV DNA transitioning to
precancer.\cite{clarke2018, hernandez-lopez2019, kelly2019} Unfortunately,
such tests are too expensive to routinely carry out for all HPV positive
women. Group testing could offer one way of making such testing feasible.

\subsection{Biomarker presence in a cohort study} In many cohort studies
specimens collected from individual participants are to be screened for a
variety of biomarkers. For example, the Connect study is a cohort study
funded by the National Cancer Institute planning to enroll 200,000 adults in
the United States with the goal of understanding the etiology of cancer
through longitudinal assessment of biomarkers, environmental exposure, and
the occurrence of cancer precursors. One biomarker of interest in this study
is monoclonal gammopathy of undetermined significance (MGUS), a premalignant
plasma cell disorder present in about 3\% of adults.\cite{landgren2019} MGUS
is a precursor for multiple myeloma and other blood cancers. Screening all
cohort participants for MGUS would add significant costs to the Connect
study, a particular concern since it is one of many biomarkers of interest.
In theory, this is an ideal case for group testing since the low prevalence
of MGUS would result in a large reduction in the number of tests to screen
the entire cohort.\cite{kyle2006} In practice, however, the sensitivity of
pooling procedures when screening for MGUS is unknown, and a validation study
would be required to characterize test performance for grouped samples and
then to assess the feasibility of group testing in this case. When considered
for this particular study, the costs of this validation study would have to
be considered in light of the, initially unknown, potential savings from a
pooling design.

\section{Notation and Dorfman procedure} For a screening program, we assume a
population of $N$ individuals for which the presence of a binary
characteristic (e.g., an infectious disease or an HIV viral load $>1000$
copies/ml) is represented by random variables $x_i,\ i = 1, 2, \ldots, N$.
When referring to a population, we mean all potentially tested individuals
with similar characteristics including demographics, type and temporality of
testing, and selection of the individual samples for testing. We will assume
that each member of this population has an identical probability, $p$, of
having said characteristic so that $x_i \sim Bernoulli(p),\ i = 1, 2, \ldots,
N$ and that each $x_i$ is independent of the others. For grouping, often a
maximum feasible group size will exist which we denote by $k_{max}$. The set
of possible group sizes is then $\mathcal{K} = \left\{1, 2, \ldots,
k_{max}\right\}$. For a group of size $k$, let $\tilde{y}^{(k)}$ be a random
variable which is 1 if at least one member of the group has the given
characteristic and 0 otherwise. Then, since each $x_i$ is independent,
$\tilde{y}^{(k)} \sim Bernoulli(1 - (1-p)^k)$.\cite{dorfman1943} Since
testing error may make observation of $\tilde{y}^{(k)}$ impossible, let
$y^{(k)}$ represent the observed value of this random variable from a test
based on a given assay. For the assay of interest, we define the sensitivity
and specificity for a grouped test containing $k$ individuals to be $Se(k) =
P(y^{(k)} = 1|\tilde{y}^{(k)} = 1)$ and $Sp(k) = P(y^{(k)} =
0|\tilde{y}^{(k)} = 0)$, respectively. We consider that these probabilities
may change with the group size, which is referred to as differential
misclassification. When the misclassification parameters are constant for any
$k$, we say that there is non-differential misclassification. Using these
definitions, the probability of a positive test is $P(y^{(k)} = 1) =
Se(k)P(\tilde{y}^{(k)} = 1) + (1 - Sp(k))P(\tilde{y}^{(k)} = 0) = Se(k) - (Se(k)
+ Sp(k) - 1) (1 - p)^k$ so that $y^{(k)} \sim Bernoulli(Se(k) - (Se(k) +
Sp(k) - 1) (1 - p)^k)$.

The first published group testing procedure proposed by
Dorfman\cite{dorfman1943} is a simple two stage procedure. To implement it, a
group size $k$ is chosen and the population is divided into groups of that size.
For each group, an initial grouped test is carried out to determine if any of
the samples within are positive for the disease. If negative, each sample is
assumed to be disease free. If positive, additional samples from each individual
in the group are tested to identify those with the disease.  If $T$ is the
number of tests required to assign a status to an individual, the choice of $k$
is typically made to minimize the expected $T$ (i.e., the expected number of
tests per person) which is given by
\begin{align*}\mathrm{E}(T| p, k, Se(k)) &= \frac{1}{k}\left[P(y^{(k)} = 0) + (k + 1) P(y^{(k)} = 1)\right]\\
  &= \frac{1}{k}\left[kP(y^{(k)} = 1) + 1\right] \\
  &= Se(k) - (Se(k) + Sp(k) - 1)(1 - p)^k + \frac{1}{k}.\end{align*}Typically, such designs are
optimized with respect to the expected number of tests.

Another important quantity which is not typically accounted for is the
overall sensitivity (specificity) of a test defined as the probability that a
positive (negative) individual is correctly identified as positive (negative)
at the termination of the pooling procedure (e.g., for a positive individual,
the probability that the test at each stage of a group in which they are a
member is positive). For clinicians and researchers, these quantities are the
most important since they allow for an understanding of how many false
positives and negatives can be expected in a population when using a group
testing procedure and have simple individualistic interpretations (since they express the
misclassification probabilities for an individual being screened).

Our reason for focusing on the Dorfman procedure here over more complicated
designs with more stages is two fold. First, it is a simple intuitive design
which can easily be implemented and explained to researchers. Second, by
requiring only two stages, the Dorfman procedure will typically maximize the
overall procedure sensitivity. To see why this is so, note that, even if
misclassification probabilities do not change with group sizes (e.g., they
are non-differential), group testing will generally lead to smaller
sensitivities and larger specificities as a result of repeated testing (e.g.,
repeated testing yields more chances for a mistake). Generally, this means
that overall sensitivity will decrease with the number of stages in a group
testing procedure.

Much of the more recent group testing
literature has focused on alternative schemes, such as those that create
smaller subgroups following a positive test prior to individual testing,
which generally have a smaller number of expected tests when compared with
the Dorfman procedure. However, each additional stage will decrease the
overall sensitivity of the screening program and result in an overall
sensitivity which is difficult to quantify. The Dorfman procedure,
however yields a simple closed form expression for the overall sensitivity.
For example, if a single unit test is treated as a gold standard (no
misclassification) the overall sensitivity of the Dorfman procedure with
initial group size $k$ will be $Se(k)$. It will be very hard, if not
impossible, for any reasonable group testing design to have a lower overall
sensitivity of the Dorfman procedure.

\section{Testing with misclassification and comparison with single testing}
Since nearly all of the issues impeding the use of group testing in medical
settings involve questions of misclassification, we briefly review the
literature related to this issue here. Beginning in the 1970s with the
resurgence of research in group testing, methodology for accounting for
misclassification was proposed. The idea is that even if a test on a single
sample has little or no misclassification, it is natural to think that there
may be measurement error induced by the combining of samples across
individuals. Graff and Roeloffs (1972)\cite{graff1972} and Hwang
(1976a)\cite{hwang1976} recognized early that the objective function to
minimize should not simply be the expected number of test when tests can be
misclassified. Graff and Roeloffs (1972)\cite{graff1972} proposed a
modification of the Dorfman procedure and searched for a design that
minimizes total cost as a linear function of the expected number of tests,
weighted expected number of good items misclassified as defective, and
weighted expected number of defective items misclassified as good. Burns et
al. (1987)\cite{burns1987} generalized Graff and Roeloffs (1972) results to
the situation where the probability of misclassification depends on the
proportion of defective items in the group.

Hwang (1976a) \cite{hwang1976}
studied a group testing model with the presence of a dilution effect, where a
group containing a few defective items may be misidentified as a group
containing no such items, especially when the size of the group is large. He
calculated the expected cost under the Dorfman procedure in the presence of
the dilution effect and derived the optimal group sizes to minimize this
cost. Further, Wein and Zenios (1996)\cite{wein1996} embedded a group testing
model for continuous test outcomes into a dynamic programming algorithm that
derives a group testing design to minimize a linear combination of expected
cost due to false negatives, false positives, and testing. Malinovsky et al.
(2016)\cite{malinovsky2016} characterized the optimal design in the Dorfman
procedure in the presence of non-differential misclassification by maximizing
the ratio between the expected number of correct classifications and the
expected number of tests. Using the same criterion and testing procedure,
they also characterized a cut-off point of disease prevalence where all
individuals should be tested together at the first stage.

Aprahamian et al.
(2019)\cite{aprahamian2019} considered the Dorfman procedure in the
population with heterogeneous prevalences\citep{hwang1975, hwang1981} under
the setting of non-differential misclassification. They investigated two
models: in the first one a linear combination of the expected number of
false positives, false negatives and total number of tests was minimized; in
the second one a linear combination of the expected number of false positives
and false negatives was minimized, subject to constraints on the upper bound
of the expected total number of tests. In contrast to earlier work, recent
authors have argued that the expected number of tests alone should be used and that
careful accounting for the number of correct classifications is an
unnecessary complication \cite{hitt2019}. The basis for Hitt et
al.'s\cite{hitt2019} argument that misclassification need not be considered
in optimal design is based on a comparison of the expected number of tests
versus the ratio of the expected number of tests and the expected number of
correct classifications. Note that the above citations do not provide an
exhaustive list of approaches and many additional works have addressed group
testing under misclassification (see, for example,\cite{zenios1998,
gupta1999, kim2007, may2010, bilder2012, liu2017})

Many recent papers assume non-differential misclassification. Specifically,
they assume that misclassification does not depend on the size of the group
and that there is misclassification for a test used for on a single or
individual sample. Although the assumption of non-differential
misclassification may be reasonable for some types of sample pooling, it
cannot be generally assumed. Further, misclassification needs to be defined
relative to a gold standard. A natural comparison of group testing screening
designs is with a design where individuals are tested separately. In many
cases, it is reasonable to assume that the assay tested on a single sample is
the gold standard. In this case, misclassification for group testing will be
relative to single testing with the sensitivity and specificity of the
individual test being 1, as it also was assumed earlier by Hwang
(1976a)\cite{hwang1976}.

\section{ What is the optimal design?}
As can be seen from the previous section, many works have appropriately attempted to account for
misclassification for the optimal group testing design for disease identification. Minimizing the expected
number of tests has been used as an objective function.\cite{hitt2019}
In other previous works, authors have proposed minimizing a linear combination of the
expected number of tests and the rate of correct classification.\citep{graff1972, burns1987,
aprahamian2019} However, choosing the coefficients for these terms is subjective and may be difficult to
motivate from  a medical or public health perspective. Some authors include a cost for incorrect
classification\citep{hwang1976, wein1996} which may also difficult to motivate from  a medical or
public health perspective.  Although, the criterion proposed by\cite{malinovsky2016} does not
require such specifications, it assigns the same weight for the expected number of tests and
the expected number of correct classifications, and therefore can also be subjective.

A larger issue is that none of these cited works have considered the impact of
differential misclassification. In this case, an optimization procedure which
does not constrain the misclassification parameters may lead to unacceptably
low values of sensitivity and/or specificity and overly optimistic estimates of
the savings provided by group testing. For example, based on the Dorfman
procedure, the expected number of tests per person (or individual) is a decreasing function of
the sensitivity. Therefore, if sensitivity is changing with group sizes, the
group size minimizing the expected number of tests may result in very low
sensitivity. In this work, we address both of these issues by proposing a
simple, easy to interpret, optimization problem so that the group size
$k$ is chosen to satisfy the following:

\begin{align}
\begin{split}
    \label{eq:opt_1}
  &\argmin_{k \in \mathcal{K}} \mathrm{E}(T|p, k, Se(k), Sp(k)), \\
  &\text{subject to}\ Se(k) \geq \delta_1, \\
  &\text{subject to}\ Sp(k) \geq \delta_2,
\end{split}
\end{align}
where $\delta_1$ and $\delta_2$ are fixed threshold values and,  $p$, $Se(k)$, and $Sp(k)$ are assumed known.
Here, the objective function is the expected number of tests per person (or individual)
and the misclassification parameters are subject to lower bounds. This has the benefit of being very easy to interpret
and explain to non-statisticians. Furthermore, it ensures that the misclassification parameters are sufficiently high
in the final design.

\section{Importance of correctly specifying Se(k) and Sp(k)} Despite the
advantages of this approach, a major drawback (which is shared with all
previous approaches), is that it relies on the assumption that the parameters
$Se(k)$ and $Sp(k)$ are known. In reality, this is almost never the case and
researchers are likely to have little a priori information on the magnitude
of the misclassification parameters and whether or not they change with the
group size. As such, to choose an optimal design and understand its
properties, it is essential that researchers first acquire knowledge of the
diagnostic performance of the assay for grouped samples. Without such
knowledge no claim that group testing is more efficient than individual
testing for disease screening can confidently be made. By far the most common
approach in the group testing literature is to use estimates of sensitivity
and specificity found in the literature. There are a number of issues of
concern here. First, the estimates of sensitivity and specificity are often based
on studies conducted in other populations. The problem of applying the
sensitivity and specificity of an assay in one population when they were
estimated in a different population with a different mix of patients has been
well recognized in the area of diagnostic medicine.\cite{ransohoff1978}
Second, the uncertainty in the estimation of sensitivity and specificity is
not taken into account in most comparisons.

Misspecification of the sensitivity and specificity can impact the testing procedure in two primary
ways. First, even small differences can lead to changes in the optimal choice of design. This is
particularly true if the sensitivity changes with the group size, since the expected number of tests
typically decreases with the sensitivity. This will result in a poor understanding of the expected
number of tests for a given design and may lead to poor decisions regarding the application of a
group testing procedure in a given population.

Second, misspecification of the misclassification parameters can lead to choosing a design with very
high error rates. For example, if the sensitivity of an assay is decreasing with group size then
overestimating the sensitivity by even a small degree can lead to choosing a large group size for which
the assumed overall error rate estimate is overly optimistic.

These issues are illustrated in the following example.

\subsection{Example}

\begin{figure}[!htb]
    \caption{Plot of sensitivity function, $Se(k) = f_H(p, k, d)$ for $p = 0.1$ for various values of $k$ and $d$.}
  \label{fig:hwang}
  \centering
 \includegraphics[width = .5\columnwidth]{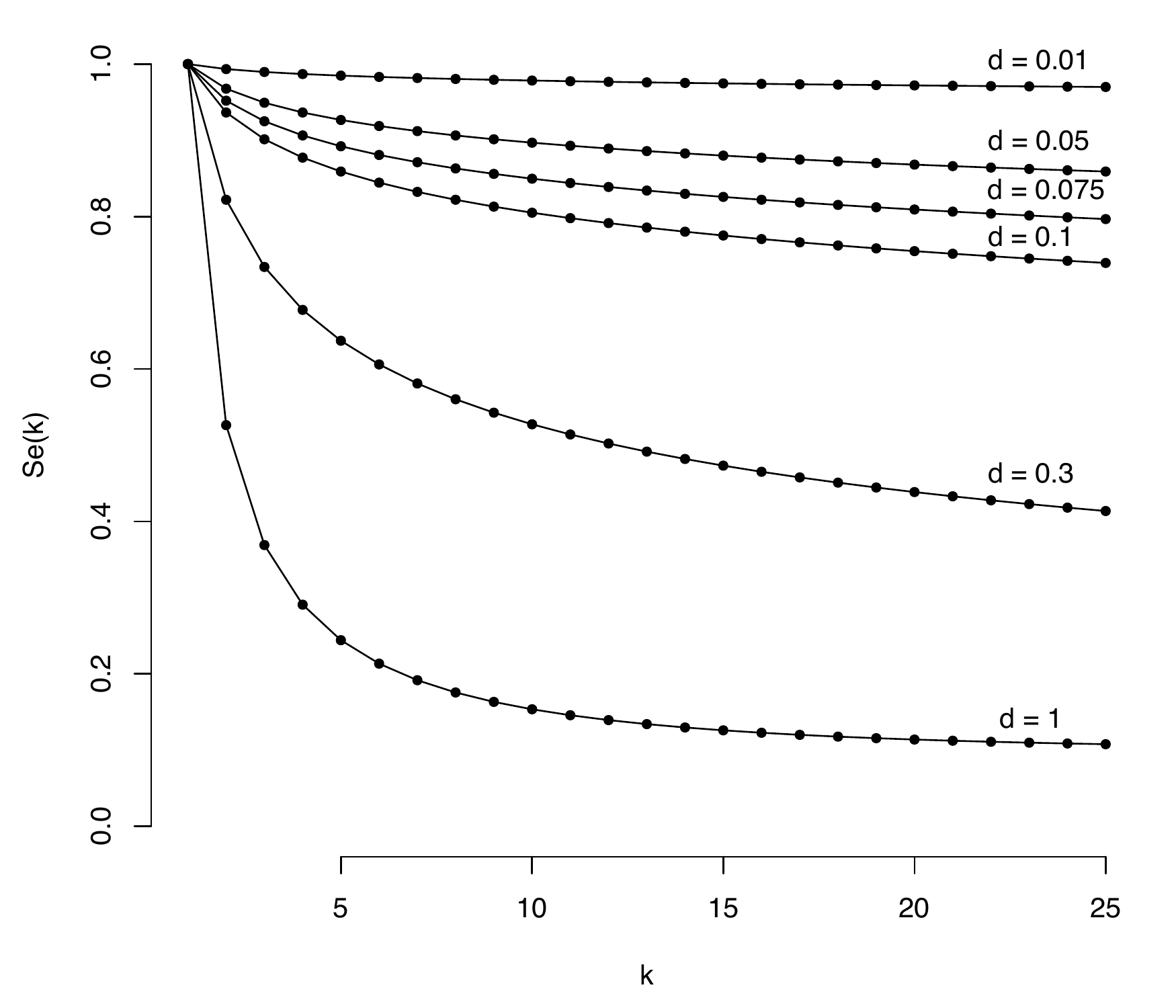}
\end{figure}

In this section, we explore numerically how misspecification of the
sensitivity function, $Se(k)$, when choosing an optimal design can lead to
errors in the estimation of a procedure's overall sensitivity as well as the
expected number of tests. We will assume in all cases that $Sp(k) = 1$. While
this assumption will not be true in many settings, it is sufficient here to
illustrate how poor a-priori information regarding the sensitivity function
can lead to bad design choices. Furthermore, grouped specificity is often of
only secondary interest to researchers as, even if the specificity decreases
with the group size, group testing still results in a larger overall
specificity than individual testing. This is true since, to ultimately be
determined positive, an individual must undergo testing at least two stages,
one of which is at the individual level. To illustrate different potential
ways the sensitivity might be subject to differential misclassification, we
consider the function $Se(k) = f_H(p, k, d) = \frac{p}{1 - (1 - p)^{k^d}}$,
due to Hwang (1976a)\cite{hwang1976}, for various values of $d$. This
function allows us to model the sensitivity as a function of the group size
$k$ using an index parameter $d$ for which the sensitivity decreases as $d$
ranges from $0$ to $1$. Note that, when $d = 0$, $f_H(p, k, 0) = 1$
indicating that the grouped assay is perfectly sensitive (i.e., false
negatives do not occur). For $d = 1$, $f_H$ gives the probability of a single
unit being positive given there is at least one positive in the group. Plots
of $f_H$ for $p = 0.1$ and $k = 1, 2, \ldots, 25$ for various values of $d$
are shown in Figure \ref{fig:hwang}. For our example, we assume that the true
assay sensitivity can be represented by taking $d = 0.075$, and will look at
designs constructed assuming values of $d = 0.01, 0.05, 0.1$, and $0.3$. The
value of $d = 0.075$ represents a moderate decay of the sensitivity function
as $k$ increases and allows for the comparison of cases when the sensitivity
function is both underestimated (e.g., $d$ is assumed to be greater than
$0.075$) and overestimated (e.g., $d$ is assumed to be less than $0.075$). In
this context, underestimation implies that we assume the sensitivity function
decreases more quickly with increasing group sizes and reaches a lower point
than is true. Likewise, overestimation implies that we assume the sensitivity
function decreases more slowly with increasing group sizes and does not
decrease as far as what is true.

To find the optimal group size, $k_{opt}$, we consider two approaches. The first, for a given $p$ and
function $Se(k)$, solves the unconstrained optimization problem
\[\argmin_{k \in \mathcal{K}} \mathrm{E}(T|p, k, Se(k)),\]
where $\mathcal{K}$ is the set of all possible group sizes and \[\mathrm{E}(T|p, k, Se(k)) =
\left\{\begin{array}{lc} 1, & k = 1 \\ Se(k)[1 - (1 - p)^k] + 1 / k, & k \geq 2 \end{array} \right.\]

The second approach enforces a lower bound on the overall sensitivity of a procedure by finding $k$
which satisfies:

\begin{align}
\begin{split}
    \label{eq:opt}
  &\argmin_{k \in \mathcal{K}} \mathrm{E}(T|p, k, Se(k)), \\
  &\text{subject to}\ Se(k) \geq \delta,
\end{split}
\end{align}
where $\delta$ is a fixed threshold value and, again, $p$ and $Se(k)$ are assumed known.

For our numerical comparisons, we set $\mathcal{K} = \{1, 2, \ldots, 25\}$ and $\delta = 0.95$.
Results are shown in Table \ref{tab:comp1}. The table contains estimates for three basic quantities:
\begin{itemize}
    \item  $k_{opt}$, the optimal  group size chosen for a given optimization procedure;
    \item $Se(k_{opt})$, the sensitivity function evaluated at the optimal group size;
    \item $\mathrm{E}(T|k_{opt})$, the expected number of tests for a given procedure based on the
        optimal group size.
\end{itemize}

For each quantity, an assumed value (e.g., the value calculated based on the
sensitivity function $Se(k) = f_H(p, k, d)$ for a given $d$) is indicated by a hat, $\hat{\cdot}$ (e.g.,
$\mathrm{E}(T|\hat{k}_{opt})$ is the true expected number of tests based on
the assumed optimal group size and true sensitivity value and
$\widehat{\mathrm{E}}(T|\hat{k}_{opt})$ is the value believed to be true
based on the assumed sensitivity values).

\begin{table}[!htb]
  \caption{Assumed design parameters when the true sensitivity function is $Se(k) = f_H(p, k, d =
    0.075)$ and $\hat{k}_{opt}$ is chosen with the assumption that $d = 0.01, 0.05, 0.1$ or $0.3$. Results are
    given for designs chosen based on solving the unconstrained and constrained optimization
    problems.}
  \label{tab:comp1}
\centering
\resizebox{\columnwidth}{!}{%
\begin{tabular}{ccccccccccccccccccc}
\toprule
  && \multicolumn8c{Unconstrained} & \multicolumn8c{Constrained}\\
  \cmidrule(lr){3-10} \cmidrule(lr){11-18}
$p$ & $d$ & $\hat{k}_{opt}$ & $k_{opt}$ & $\widehat{\mathrm{Se}}(\hat{k}_{opt})$ & $\mathrm{Se}(\hat{k}_{opt})$ & $\mathrm{Se}(k_{opt})$ &
$\widehat{\mathrm{E}}(T|\hat{k}_{opt})$ & $\mathrm{E}(T|\hat{k}_{opt})$ & $\mathrm{E}(T|k_{opt})$& $\hat{k}_{opt}$ & $k_{opt}$ & $\widehat{\mathrm{Se}}(\hat{k}_{opt})$ & $\mathrm{Se}(\hat{k}_{opt})$ & $\mathrm{Se}(k_{opt})$ &
$\widehat{\mathrm{E}}(T|\hat{k}_{opt})$ & $\mathrm{E}(T|\hat{k}_{opt})$ & $\mathrm{E}(T|k_{opt})$\\
 \cmidrule(lr){1-1}   \cmidrule(lr){2-2}   \cmidrule(lr){3-3}   \cmidrule(lr){4-4}  
                                                                                     \cmidrule(lr){5-5}   \cmidrule(lr){6-6}   \cmidrule(lr){7-7}   \cmidrule(lr){8-8}   \cmidrule(lr){9-9}   \cmidrule(lr){10-10}   \cmidrule(lr){11-11}   \cmidrule(lr){12-12}   \cmidrule(lr){13-13}   \cmidrule(lr){14-14}   \cmidrule(lr){15-15}   \cmidrule(lr){16-16}   \cmidrule(lr){17-17}   \cmidrule(lr){18-18}

0.01   &  0   &  11   &  12   &  1   &  0.836   &  0.831   &  0.196   &  0.178   &  0.178   &  11   &  1   &  1   &  0.836   &  1   &  0.196   &  0.178   &  1    \\
   &  0.01   &  11   &  12   &  0.976   &  0.836   &  0.831   &  0.193   &  0.178   &  0.178   &  11   &  1   &  0.976   &  0.836   &  1   &  0.193   &  0.178   &  1    \\
   &  0.05   &  12   &  12   &  0.884   &  0.831   &  0.831   &  0.184   &  0.178   &  0.178   &  2   &  1   &  0.966   &  0.95   &  1   &  0.519   &  0.519   &  1    \\
   &  0.1   &  13   &  12   &  0.775   &  0.826   &  0.831   &  0.172   &  0.178   &  0.178   &  1   &  1   &  1   &  1   &  1   &  1   &  1   &  1    \\
   &  0.3   &  21   &  12   &  0.404   &  0.797   &  0.831   &  0.125   &  0.199   &  0.178   &  1   &  1   &  1   &  1   &  1   &  1   &  1   &  1    \\
0.05   &  0   &  5   &  6   &  1   &  0.889   &  0.877   &  0.426   &  0.401   &  0.399   &  5   &  2   &  1   &  0.889   &  0.951   &  0.426   &  0.401   &  0.593    \\
   & 0.01   &  5   &  6   &  0.984   &  0.889   &  0.877   &  0.423   &  0.401   &  0.399   &  5   &  2   &  0.984   &  0.889   &  0.951   &  0.423   &  0.401   &  0.593    \\
   & 0.05   &  5   &  6   &  0.925   &  0.889   &  0.877   &  0.409   &  0.401   &  0.399   &  2   &  2   &  0.967   &  0.951   &  0.951   &  0.594   &  0.593   &  0.593    \\
   & 0.1   &  6   &  6   &  0.84   &  0.877   &  0.877   &  0.389   &  0.399   &  0.399   &  1   &  2   &  1   &  1   &  0.951   &  1   &  1   &  0.593    \\
   &  0.3   &  10   &  6   &  0.514   &  0.845   &  0.877   &  0.306   &  0.439   &  0.399   &  1   &  2   &  1   &  1   &  0.951   &  1   &  1   &  0.593    \\
0.1    &  0   &  4   &  4   &  1   &  0.906   &  0.906   &  0.594   &  0.562   &  0.562   &  4   &  2   &  1   &  0.906   &  0.952   &  0.594   &  0.562   &  0.681    \\
   &  0.01   &  4   &  4   &  0.987   &  0.906   &  0.906   &  0.589   &  0.562   &  0.562   &  4   &  2   &  0.987   &  0.906   &  0.952   &  0.589   &  0.562   &  0.681    \\
   &  0.05   &  4   &  4   &  0.937   &  0.906   &  0.906   &  0.572   &  0.562   &  0.562   &  2   &  2   &  0.968   &  0.952   &  0.952   &  0.684   &  0.681   &  0.681    \\
   &  0.1   &  4   &  4   &  0.877   &  0.906   &  0.906   &  0.552   &  0.562   &  0.562   &  1   &  2   &  1   &  1   &  0.952   &  1   &  1   &  0.681    \\
   &  0.3   &  25   &  4   &  0.414   &  0.797   &  0.906   &  0.424   &  0.779   &  0.562   &  1   &  2   &  1   &  1   &  0.952   &  1   &  1   &  0.681    \\
\bottomrule
\end{tabular}
}
\end{table}

From the results, we see that the unconstrained optimization, which considers
only the expected number of tests, often yields very poor assumed overall
sensitivity. For examples, with $p=0.01$ and $d$ assumed to be $0.01$, the
assumed sensitivity value is 0.976, which is substantially different from the
actual sensitivity value of $0.836$ which would occur if $d=0.075$ and $k=11$
were used. This highlights the fact that such an optimization procedure can do
nothing to control the overall sensitivity rates and should be used cautiously,
particularly with differential misclassification. While the overall sensitivity
values are always much higher when using the constrained procedure, when the
assumed sensitivity function overestimates the true values the chosen group
size can yield an overall sensitivity value much smaller than is assumed to be
true. This can be seen by, again, looking at the example of $p=0.01$ with $d$
assumed to be $0.01$ where the assumed and actual sensitivity values are
$0.976$ and $0.836$, respectively. This highlights a major drawback of the
constrained procedure (which is also present in the unconstrained), namely
that it can not overcome poor a-priori information concerning the sensitivity
function (specifically, overestimation of the sensitivity function).

The differences observed between true and assumed sensitivity tend to
decrease sharply as $p$ increases. For example, looking again at an assumed
value of $d=0.01$, we noted above that the assumed and true sensitivity
values were $0.976$ and $0.836$, respectively, when $p=0.01$. At $p=0.1$,
these values have become $0.987$ and $0.906$. This is due to the fact that
for larger $p$ the expected number of tests decreases rapidly with the group
size, regardless of the sensitivity values. Differences between the assumed
and true expected number of tests, however, follows the opposite pattern,
with the differences increasing with $p$. For example, with an assumed value
of $d = 0.3$, the difference between the true and assumed expected number of
tests per person (or individual) increases from $0.199-0.125 = 0.074$ at $p =
0.01$ to $0.779 - 0.424 = 0.355$ at $p = 0.1$. Such a difference would lead
to an overconfidence in the savings provided by group testing which scales
linearly with the total number of people to be screened. For the most part,
we see that, when the sensitivity function is assumed to have lower values
than are true for each group size, the expected number of tests per person is
underestimated (as seen in the previous example). Conversely, when the
sensitivity function is assumed to have higher values than are true for each
group size, the expected number of tests per person (or individual)is
overestimated. For example, returning to the case of $p=0.01$ and an assumed
value of $d=0.01$, the assumed expected number of tests exceeds the true
expected number of tests by $0.193 - 0.178 = 0.015$.

\section{Optimal design incorporating estimation error in $\widehat{Se(k)}$
and $\widehat{Sp(k)}$} As seen in the previous example, reliable knowledge of
the misclassification parameters and their dependence on $k$ is essential in
designing a group testing screening program. In most cases, this will require
researchers to first obtain population specific estimates of the sensitivity
and specificity. To do this will require a validation study design in which
individuals with known disease status (most likely from initial individual
screening) are tested in groups of varying group sizes. To date, we are
unaware of any literature related to the question of how to best design such
studies. However, in practice it is important to consider how large such
validation studies would need to be before deciding if group testing is a
reasonable approach. Another important question is, given that a validation
study of a certain size is to be carried out, how large of a target
population for screening is required to see an overall benefit from utilizing
group testing. Answers to such questions will vary greatly depending on the
underlying population and particular assay being used, but it is reasonable
to assume that such considerations will show group testing is not warranted
in many situations when such an approach might otherwise be considered.

\subsection{Example} To estimate $Se(k)$ and $Sp(k)$, a simple validation design
is described in Algorithm \ref{alg:val} for an initial sample of size $N$ and a
maximum group size $k_{max}$. The maximum group size is predetermined by
researchers to be the largest possible group size under consideration. Once the
misclassification parameters have been estimated, they can be used to find $k_{opt}$
from the constrained optimization procedure described above in (\ref{eq:opt})
using the estimated sensitivity and specificity values. In this section, our
goal is to determine how large of an initial validation sample size, $N$, would
be required to be confident that the bounding criterion in (\ref{eq:opt}) is
truly met. Mathematically, for the sensitivity this means we hope to achieve
$\phi(\delta) = \mathrm{P}(\widehat{Se(\hat{k}_{opt})} > \delta) > \epsilon$
where $\epsilon$ is some threshold value. Note that this is conceptually similar to a
tolerance interval where we can be assured with a certain probability that a
particular value falls within the interval.

To determine the necessary validation size, $N$, we conducted a simulation
study with 50,000 simulations and found the smallest $N$ such that the
empirical probability $\phi(0.95) > 0.95$. The full simulation algorithm is
described in Algorithm \ref{alg:pseudo}. As above, we assumed $Sp(k) = 1$ for
all $k$. For the sensitivity functions, we considered several possibilities:
\[Se_1(k) = 1 - 0.02(k - 1),\] \[Se_2(k) = \begin{cases}1 - 0.02\times 2^{k /
2}, & k = 1, 2, \ldots, 11\\ 0, & \text{otherwise} \end{cases}\] \[Se_3(k) =
f_H(p, k, d = 0.1),\] \[Se_4(k) = f_H(p, k, d = 0.3).\]

Simulations were carried out for $p = 0.01, 0.02, \ldots, 0.10$ and with $k_{max} = 10$.

\begin{figure}[!htb]
  \caption{Barplots showing required validation study size, $N$, total number of assay tests in
    validation study, $T_V$, and minimum population size to see a benefit from group testing, $N^*$
    for various underlying true sensitivity functions. The bottom axes are values of $p$.}
  \label{fig:N}
  \includegraphics[width = \columnwidth]{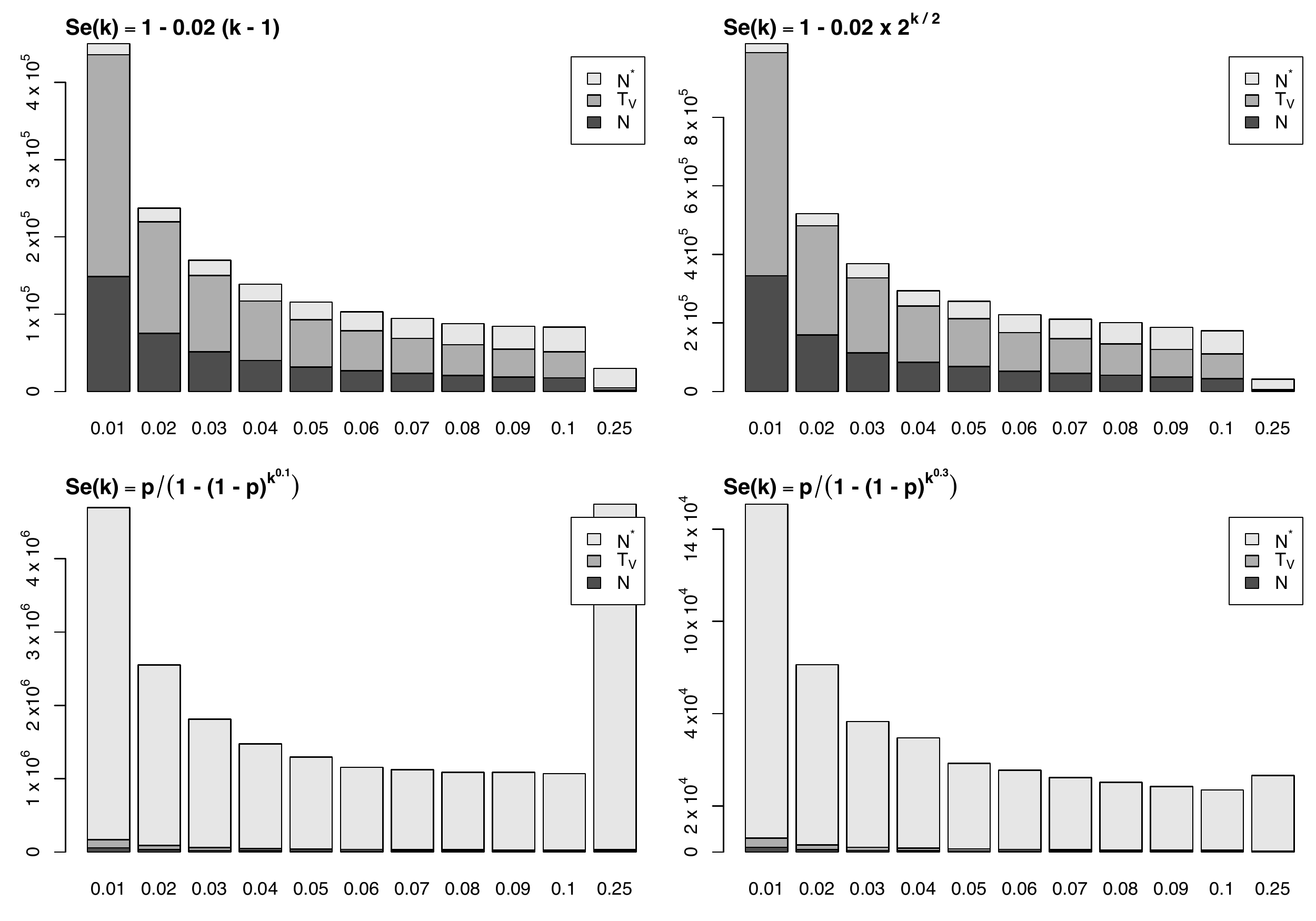}
\end{figure}

Once the smallest $N$ was determined, we found the smallest total population
size, $N^*$, required to see a benefit from group testing following such a
validation procedure. This value was determined by comparing the expected
number of tests required to complete screening the population plus the total
number of tests used in the validation study to the total population size,
which represents the number of tests required under individual testing. Letting $T_V$ be the total number of tests
required in the validation study, $N^*$ can be found by solving the inequality \[(N^* - N) E(T| p, k, Se(k)) +
T_V \leq N^*,\] or equivalently \[\frac{T_V- N\times E(T| p, k, Se(k))}{1 -
E(T| p, k, Se(k))} \leq N^*.\] The
expected value in this expression was taken as the average expected value
across all simulations for the given validation sample size. Results are
shown in Figure \ref{fig:N}.

\begin{algorithm}[!htb]
  \caption{Procedure for validation study.}
  \label{alg:val}
\begin{algorithmic}
  \For{$k = 1, 2, \ldots, k_{max}$}
    \If{$N / k$ is an integer}
      Randomly group units into $N / k$ groups of size $k$\;
    \Else~
    Randomly form $\left\lfloor{N / k}\right \rfloor$
    groups of size $k$ and construct a final group with the remaining $N - k~\times \left\lfloor{N/k}\right \rfloor$ units
    and $k - N + k~\times \left\lfloor{N/k}\right \rfloor$ duplicate units randomly chosen from the other
    groups\;
    \EndIf
    \If{$k > 1$}
    Assuming estimates from the $k=1$ stage are correct, estimate $Se(k) = \frac{\# \text{groups with at least one positive member and
        testing positive}}{\# \text{groups with at least one positive member}}$ and $Sp(k) = \frac{\# \text{groups with no positive members and
        testing negative}}{\# \text{groups with no positive members}}$
    \EndIf
\EndFor
\end{algorithmic}
\end{algorithm}

\begin{algorithm}[!htb]
  \caption{Pseudo-code for validation study simulations.}
  \label{alg:pseudo}
  \begin{algorithmic}
  \State{Let $N = 10,000$\;}
  \State{Let $N_{max} = 0$\;}
  \State{Let $N_{min} = 0$\;}
  \While{true}
  \For{$i = 1, 2, \ldots, 50,000$}
    \State{Calculate $\widehat{Se(k)}_i$ and $\widehat{Sp(k)}_i$ using Algorithm \ref{alg:val} for each $k = 1, 2, \ldots, 10$\;
    Find $\hat{k}_{opt}$ solving (\ref{eq:opt}) with $\delta = 0.95$ based on these estimates\;}
    \State{Set $\psi_i = \begin{cases} 1, & \widehat{Se(\hat{k}_{opt})}_i > 0.95\\ 0, & \text{otherwise}\;\end{cases} $}
    \EndFor
    \State{Calculate $\hat{\phi}(0.95) = \sum_{i = 1}^{50,000} \frac{\psi_i}{50000}$}
    \If{$\hat{\phi}(0.95) - 0.95 > 0.01$}
    Let $N_{max} = N$\;
    Let $N = \left\lfloor{\frac{N + N_{min}}{2}}\right \rfloor$\;

    \ElsIf{$\hat{\phi}(0.95) - 0.95 < -0.01$}
    Let $N_{min} = N$\;
    \If{$N_{max} = 0$}
        Let $N = 2N$\;
        \Else
        Let $N = \left\lfloor{\frac{N + N_{max}}{2}}\right \rfloor$\;
      \EndIf
    \Else~
    exit while loop
      \EndIf
      \EndWhile
  \end{algorithmic}
\end{algorithm}

Unsurprisingly, for all sensitivity functions we see the required validation sample sizes decrease
with increasing prevalence. This is expected as smaller numbers of individuals are required to
ensure an adequate number of groups with at least one positive member. For the sensitivity functions
on the top row, the decrease in sensitivity is more gradual so that a larger group size can be
chosen. For these cases, larger validation sample sizes are required to accurately estimate the
sensitivity function. However, since the larger group sizes will allow for a smaller expected number
of tests, the additional sample size required to see a benefit from group testing is small.

\section{Revisiting the case studies} \subsection{Population based screening
for SARS-CoV-2 infection} Large scale screening for SARS-CoV-2 is an important
and pressing public health issue. Implementation of group testing to
facilitate such screening currently faces several obstacles which must be
considered before beginning such a program. First, prevalence values in a
given region are unknown and constantly changing. This forces any design
choices to be made somewhat ad hoc. This is particularly an issue as testing
protocols and indications are currently in flux across state and national
health departments, so that the underlying screening population
characteristics can change at any time. For example, if the positive rate of
individual tests in an area is approximately 5\%, a procedure testing groups
of size 5 could offer significant advantages. If, however, at a later time
the prevalence of the screening population increases to 29\%, or greater, due
to testing only individuals at higher risk such a procedure would require
more tests than individual screening. The inability of testing facilities to
anticipate such swings could lead to very expensive mistakes.

A second issue is that it is not known a priori how assay sensitivity changes with group sizes.
This is particularly true as such values may vary across populations and labs given the wide range
of testing techniques currently being implemented. While a validation study would be feasible for
such a use case, it is unlikely that public health officials would be willing to reallocate sparse
testing materials for large speculative studies at this time. In a public health crisis like
COVID-19, we recommend that samples be stored so that, at a minimum, the feasibility of group
testing can be evaluated at a later time.

If despite these concerns a group testing program were to be implemented, a basic prevalence estimate
could be obtained using recent individual testing data. To carry out a validation procedure as
described above we show the estimated validation sample sizes, total number of tests, and population
size required to see a benefit from group testing for prevalence values $0.05, 0.1,$ and $0.25$ in
Table \ref{tab:covid}. Values are reported based on two assumed underlying sensitivity functions: 1)
a linear function, $Se(k) = 1 - 0.02(k-1)$ and 2) the Hwang function with $d = 0.1$, $f_H(d = 0.1)$.

\begin{table}
    \caption{Estimated validation sample size, $N$, number of validation tests, $T_V$, and necessary
    population size to see a benefit from group testing, $N^*$, for a $COVID-19$ screening program
    based on prevalence values of $0.05, 0.1$, and $0.25$. Values are calculated separately for two
    underlying sensitivity functions: Se(k) = 1 - 0.02(k-1) and $f_H(d = 0.1)$.}
    \label{tab:covid}
    \centering
    \begin{tabular}{ccccccc}
        \toprule
        &\multicolumn3c{Se(k) = 1 - 0.02(k-1)} & \multicolumn3c{$f_H(d = 0.1)$}\\
        \cmidrule(lr){2-4} \cmidrule(lr){5-7}
    Prevalence & N & $T_V$ & $N^*$ & N & $T_V$ & $N^*$\\
        \cmidrule(lr){1-1} \cmidrule(lr){2-2} \cmidrule(lr){3-3}\cmidrule(lr){4-4}\cmidrule(lr){5-5}\cmidrule(lr){6-6}\cmidrule(lr){7-7}
        0.05 & 31,679 & 92,789 & 147,238 & 13,710 & 40,158 & 1,307,444 \\
        0.1 & 17,500 & 51,259 & 100,771 & 8,906 & 26,090 & 1,077,732 \\
        0.25 & 1,640 & 4,806 & 31,494 & 10,781 & 31,582 & 4,754,298 \\
        \bottomrule
    \end{tabular}
\end{table}

\subsection{Large scale screening for HIV viral load}
The specifics of designing a screening program for monitoring HIV viral load will vary as different regions
employ differing testing protocols and thresholds. As an example, we consider a case with
suspected ART failure prevalence around 9\%, a value reported among those using ART for at least 18
months for a Malawian cohort in Nicholas et al. (2019).\cite{nicholas2019} While studies evaluating
the pooled sensitivity for fixed group sizes have been done\cite{elbouzidi2014, tilghman2011}, such
values are likely cohort specific and would need to be re-estimated before application to a
specific population. Furthermore, in order to make informed decisions about an optimal group size it
would be necessary to first understand how the sensitivity changes with the group size. Using the
procedure outlined above, we can look at the sample sizes required under different assumed
sensitivity functions to evaluate how feasible group testing would be in this case. For example, if
we assumed the linear sensitivity function $Se(k) = 1 - 0.02 (k - 1)$ for a prevalence of $0.09$ we
would require 18,750 individuals to enroll in a validation study requiring 54,920 total tests and a
population size to 103,097 to see a benefit from group testing. If, however, the sensitivity
function was the Hwang function with $d = 0.1$, we would require 9,375 individuals to enroll in a
validation study requiring 27,462 tests and a population size of 1,094,884. In either case, for a
population of 600,000 screened semi-annually there is a clear potential for savings from group
testing, even after carrying out a validation procedure. Without any prior knowledge of the
sensitivity function, it would be difficult to choose an initial validation sample size as it is
impossible to give conservative bounds. Still, if the resources are available for an initial large
investment for a validation study, and health officials are able to deal with the possibility that
the pooled sensitivity will be too low for practical use, the long term and ongoing nature of HIV
viral load screening can potentially benefit largely from group testing.

\subsection{Stratified cancer screening} For HPV methylation screening, we consider a program aimed
at screening the entire US population for HPV related cervical cancer risk. This could be achieved
by collecting samples from all women, identifying those with high risk HPV subtypes (i.e., those
that act as cancer precursors), and finally administering methylation testing for the high risk HPV
group. Those with positive methylation tests would then be followed more intensely to ascertain
cervical cancer risk.  Using 2010 population estimates and estimated prevalences from
2014\cite{mcquillan2017}, we could approximate that around 20 million women in the US would test positive for a high
risk HPV subtype and we would like to design a group testing procedure to screen each of these women
using methylation testing. To date, there are no population based estimates of methylation positive
testing rates so we will assume a value of 5\% for this example. Using these values and the
validation procedure outlined above, if the underlying sensitivity function were the linear function
$Se(k) = 1 - 0.02 (k-1)$ then we would require a validation sample size of 31,679 and a total of
92,789 tests with a required population size to see a benefit from group testing of 147,238. If,
however, the true sensitivity function were the Hwang function with $d = 0.1$, we would require
13,710 women for a validation procedure requiring 40,158 tests and a total population size of
1,307,444 to see a benefit from group testing. In either case, given the large population required
for screening, group testing would likely provide large savings in this setting, even with a
necessarily large validation study. This would be true even if the actual rate of positive
methylation tests in the high risk HPV infected population were much higher. Here, the only
real impediment to using group testing would be if health officials were unwilling to accept any
additional loss of sensitivity due to pooling.

\subsection{Biomarker presence in cohort study}
For MGUS screening, we assume a prevalence of 3\% and that we would like to determine the status of
approximately 200,000 individuals. If the true sensitivity function were the linear function $Se(k) = 1 - 0.02 (k - 1)$
then we would require a sample size of 51,250 individuals for a validation procedure requiring
150,113 tests and a population size of 221,091 to see a benefit from group testing. If, however, the
true sensitivity function were the Hwang function with $d = 0.1$, we would require 21,093
individuals for a validation procedure requiring 61,785 tests and a population size of 1,832,663 to
see a benefit of group testing. Given these numbers, and lacking any a priori information on the
sensitivity function, it is unlikely that researchers would attempt to implement such a validation
procedure in this case. While the large sample sizes are offset somewhat by the need for repeat
testing, the non-trivial possibility of finding that pooling of any size reduces the sensitivity to
an unacceptable level make this an unlikely gamble for resource allocation.

\section{Discussion}
In this paper we have reviewed several of the issues faced by practitioners when deciding if group
testing can provide a feasible solution for their screening program. In this context we have
explored several issues numerically based a simple algorithm (the Dorfman two-stage procedure) and
several simplifying assumptions. In practice, there exist many additional considerations which may
alter the final decision concerning whether to implement group testing.

For all numerical comparisons, we have assumed grouping does not impact
specificity (i.e., $Sp(k) = 1$ for all $k$). While this may be reasonable in
some settings, the failure of this assumption can result in large increases
in the number of individuals required for a validation sample. In particular,
by using a minimum threshold to determine estimation accuracy we have had to
assume that $\phi(\delta)$ is monotone as a function of the validation sample
size. While this holds for $Sp(k) = 1$, this may not be true otherwise,
requiring more complicated evaluation criteria and larger sample sizes.
Furthermore, poor assumptions about $Sp(k)$ can contribute to poor estimation
of the expected number of tests and, hence, exacerbate the issues of
selecting an appropriate group testing design.

When designing our validation procedure, we made the assumption that the sensitivity does not depend
on the number of positives in a given group (i.e., we have assumed that the sensitivity is only a
function of whether or not any group member has the disease, not the full distributional makeup of the group).
In practice, this assumption may fail resulting in
significantly more complicated sensitivity functions (which must now be a function of both the group
size and the number of positives in the group). This could especially be an issue when the test
classification is a function of underlying continuous test output. If such issues could reasonably
be suspected, it would be necessary to design the validation study which accounts for this issue.

One assumption we have made is that there is a complete lack of a-priori information on the
underlying sensitivity function, necessitating the validation design to be non-parametric. However,
in cases where researchers are able/willing to make certain simplifying assumptions (e.g., that
sensitivity is linear in $k$) more efficient validation designs may be possible. In such cases,
smaller validation studies could potentially make group testing feasible in a wider range of
settings. However, given that the properties of the final design are sensitive to the correct
specification of the sensitivity function, we generally make the recommendation of a non-parametric
approach when designing important screening programs using group testing. Furthermore, if assumptions
such as monotonicity of the sensitivity as a function of group size are made, more efficient
adaptive algorithms could possibly be developed. This is an important area for future work.

We have emphasized the importance of estimating the sensitivity and specificity for different size
groups in the same population that we intend to screen. The validation study design assumes that
sample is collected from  a random sample of individuals from the population at hand and groups of
varying size be randomly formed from these samples. There are different alternative designs for the
validation sample that may lead to efficiency gains in some situations. For example, if a researcher assumes
that the specificity is 1 for all group sizes (here, we assumed it was necessary to estimate these
specificities in order to confirm this in our calculations), we may save resources by never grouping
all negative samples together.  Alternatively, rather than attempt to find the optimal design, we
could simply evaluate the properties of a group testing design for a single fixed group size. If the
false negative rate is too high, we could sequentially evaluate the properties for a smaller group
size. This approach may be advantageous for the COVID-19 example, where it is more important to
obtain a good design quickly than to spend more time to find the optimal design (i.e., the perfect
is the enemy of the good). In many cases, obtaining a random sample from a population to do a full
validation procedure may be impossible but researchers may still be interested in studying the
properties of a group testing procedure. An approach that could be taken in such cases is to use
a spiked procedure in which known concentrations of the agent being tested are included in samples
of different sizes to simulate the conditions observable in the full population. Such a procedure
could greatly reduce the number of validation tests required and could utilize pre-selected
samples, resulting in savings of both time and cost. The primary concern with such a procedure
is that the validity of the results relies heavily on the correctness of the assumptions made
concerning the underlying population characteristics and their relationship to the trait being
screened for. For a given application, researchers would have to balance their comfort level with
such assumptions with the need for empirically verified estimates.

An additional assumption we have made is that the underlying population is homogeneous with respect
to the primary trait of interest. In many cases, this is reasonable as long as the validation sample
is chosen representatively across the entire population and the subsequent samples are not grouped
based on underlying heterogeneous clusters. The impact of heterogeneity will include
additional challenges to determine the size of the validation sample and to ensure a feasible
solution to the optimization problem (\ref{eq:opt}). The issue is that even under the perfect assay setting, we need
to determine not only group sizes but also the members of the groups, and number of such
possibilities (number of the partition of the population) is astronomical even for the small
population size. In fact, under error-free testing, the optimal partition is known only for the
Dorfman procedure \citep{hwang1975, hwang1981}. From a the practical perspective, Hwang's method can be used
for Dorfman’s procedure, and the methods developed in \cite{malinovsky2019, malinovsky2020} can be used for other
group testing procedures. Another possibility, which also may be logistically easier to implement,
is a stratification of the population, such that in each stratum, there is a homogeneous population.
In such a case, the methodology developed in the present work can be used with respect to each
stratum separately.

In this paper, we have focused exclusively on the Dorfman design. In the case of a homogeneous
population, there are more efficient designs than Dorfman's two-stage procedure.\cite{sterrett1957,
sobel1959, hwang1976a} In many of these designs, the expected number of tests $E(T|p, k, Se(k))$ is not given
in closed-form, but rather calculated using recursion or dynamic programming.\cite{malinovsky2019}
In the presence of differential misclassification or dilution effects, expressions for the
expected number of tests (an important component in the objective function to evaluate) are
difficult to obtain in these cases.

Our work focused on the screening of a single disease. However, occasionally screening for multiple
diseases from a single assay may be of interest. Group testing for disease screening for multiple
diseases with test misclassification is an area for future research. With respect to feasibility, the
subject of the current paper, we want to emphasize that any design would need a validation sample
sized to be sufficient to estimate the more complex misclassification structure that would be
required for such designs.

\section{Acknowledgements}
We thank the Editor, Associate Editor, and five reviewers for their helpful comments and suggestions. The authors GH and PSA were
supported by the Intramural Program at the National Cancer Institute.

\end{document}